SolarPhys (2020) ..............
DOI10.1007/...............................# Sun-to-Earth Observations and Characteristics of Isolated Earth-Impacting Interplanetary Coronal Mass Ejections During 2008 - 2014

D. Maričić[1] · B. Vršnak[2] · A. M. Veronig[3] · M. Dumbović[2] · F. Šterc[1] · D. Roša[1] · M. Karlica[4] · D. Hržina[1] · I. Romštajn[1]© Springer ••••**Abstract** A sample of isolated Earth-impacting interplanetary coronal mass ejections (ICMEs) that occurred in the period January 2008 to August 2014 is analysed in order to study in detail the ICME *in situ* signatures with respect to the type of filament eruption related to the corresponding CME. Observations from different vantage points provided by SOHO, STEREO-A and STEREO–B are used to determine for each CME under study whether it is Earth directed or not. For Earth-directed CMEs, a kinematical study was performed using the STEREO-A,B COR1 and COR2 coronagraphs and the Heliospheric Imagers HI1, to estimate the CME arrival time at 1 AU and to link the CMEs with the corresponding *in situ* solar wind counterparts. Based on the extrapolated CME kinematics, we identified interacting CMEs, which were excluded from further analysis. Applying this approach, a set of 31 isolated Earth–impacting CMEs was unambiguously identified and related to the *in situ* measurements recorded by the Wind spacecraft. We classified the events into subsets with respect to the CME source location as well as with respect to the type of the associated filament eruption. Hence, the events are divided into three subsamples: active region (AR) CMEs, disappearing filament (DSF) CMEs, and stealthy CMEs. The related three groups of ICMEs were further divided into two subsets: magnetic obstacle (MO) events (out of which four were stealthy), covering ICMEs that at least partly expose characteristics of flux ropes, and ejecta (EJ) events, not showing such characteristics. In this way, 14 MO-ICMEs and 17 EJ-ICMES were identified. The solar source regions of the non-stealthy MO-ICMEs are found to be located predominantly (9/10; 90%) within ±30° from the solar central meridian, whereas EJ-ICMEs originate predominantly (16/17; 94%) from source regions that are outside ±30°. In the next step, MO-events were analysed in more detail, considering the magnetic field strengths and the plasma characteristics in three different segments of the ICMEs, defined as the turbulent sheath (TS), the frontal region (FR), and the MO itself. The analysis revealed various well-defined correlations for AR, DSF, and stealthy ICMEs, which we interpreted considering basic physical concepts. Our results support the hypothesis that ICMEs show different signatures depending on the *in situ* spacecraft trajectory, in terms of apex versus flank hits.Darije Maričić
dmaricic@zvjezdarnica.hr

[1] Astronomical Observatory Zagreb, Opatička 22, HR-10000 Zagreb, Croatia
[2] Hvar Observatory, Faculty of Geodesy, University of Zagreb, Zagreb, Croatia
[3] Institute of Physics & Kanzelhöhe Observatory for Solar and Environmental Research, University of Graz, Graz, Austria
[4] Sapienza University of Rome, Italia**Keywords** Coronal mass ejection · Solar wind · Filament## 1. Introduction

Coronal mass ejections (CMEs) are dynamical events in which plasma and unstable magnetic field structure are ejected from the solar corona into interplanetary space (*e.g.*, Gopalswamy, 2004). When CMEs arrive at Earth, they are detected *in situ* as interplanetary CMEs (ICMEs). On their route through the heliosphere,

SOLA: Maricic2020_v13.tex; 10 August 2020; 9:33; p. 1

ICMEs can interact with high speed solar wind streams or other ICMEs, forming complex magneto-plasmatic structures.

Current knowledge of the propagation, evolution and morphology of CMEs/ICMEs has been obtained from white light coronagraph images and *in situ* data (*e.g.*, Hundhausen *et al.*, 1984; Lepping *et al.*, 1995; Ogilivie *et al.*, 1995; Smith *et al.*, 1998; McComas *et al.*, 1998; Vourlidas *et al.*, 2010; for a recent review see Manchester *et al.*, 2017). Remote-sensing observations of CMEs by coronagraphs such as the Large Angle and Spectrometric Coronagraph (LASCO; Brueckner *et al.*, 1995) onboard the Solar and Heliospheric Observatory (SOHO) and coronagraphs and Heliospheric Imagers of the Sun–Earth Connection Coronal and Heliospheric Investigation (SECCHI) (Howard *et al.*, 2008) onboard the Solar Terrestrial Relations Observatory (STEREO) enabled imaging of CMEs from their liftoff in the lower corona up to the distance of Earth and beyond, providing insights into the CME/ICME morphology and kinematics close to the Sun as well as in the interplanetary space. Soon after the discovery of CMEs, the connection between white-light CMEs and *in situ* disturbances was recognized (*e.g.*, Sheeley *et al.*, 1980). Linking the remote CME observations with the structure of their *in situ* counterparts is not a straightforward task. First, because white light images of CMEs present only a projected view and line-of-sight integrated emission of an optically thin plasma structure, whereas the *in situ* solar wind (SW) measurements provide only local characteristics as the spacecraft passes through a large-scale ICME structure. In addition, the correct pairing of a CME imaged close to the Sun and an *in situ* observed ICME is often very difficult, especially in periods of high activity due to multiple CMEs launched in close succession.

Generally, it is assumed that all CMEs should have a flux rope structure, but *in situ* measurements indicate that on average only one-third of the identified CMEs show a clear flux rope signature, *i.e.*, only about one third of ICMEs are so-called magnetic clouds (*e.g.*, Gosling, 1990; Nakwacki *et al.*, 2011; Howard and DeForest *et al.*, 2012), representing magneto-plasma structures with increased, smoothly rotating magnetic field (Burlaga *et al.*, 1981; Zurbuchen & Richardson, 2006; Kilpua *et al.*, 2017, and references therein). However, there is no specific solar wind (SW) signature which unambiguously represents all ICMEs. Rather, we observe a subset of signatures in which some characteristics may appear and disappear during the passage of an ICME (Gosling, 1990).

Considering *in situ* signatures at 1 AU (Burlaga *et al.* 1981), ICMEs can be classified into two subsets: magnetic clouds (MC) and ejecta (EJ). According to Burlaga *et al.* (1981), MCs are well structured ICMEs, which have a magnetic field magnitude higher than average and are characterized by low magnetic field variance and smoothly rotating direction during the passage of the ICME over the *in situ* spacecraft (indicative of a flux rope), as well as by low proton temperature and low plasma-$\beta$, *i.e.* low gas-to-magnetic pressure ratio (Klein and Burlaga, 1981). When a smoothly rotating magnetic field is not observed, such an ICME is classified as an EJ (Burlaga *et al.*, 2001). However, many ICMEs contain partially an MC and partially complex structures, which could be a result of the passage of an ICME far from its apex, or a distortion of the ICME internal magnetic structure during its interplanetary evolution (see e.g. Jian *et al.*, 2006; Richardson & Cane, 2010; Kilpua *et al.*, 2017; Manchester *et al.*, 2017 and references therein). According to Jian *et al.* (2006) and more recent work of Nieves-Chinchilla *et al.* (2018, 2019), the MC definition can be extended by introducing the concept of a magnetic obstacle (MO), where MO configurations are not always defined by a full single magnetic field rotation and do not necessarily show a MC configuration throughout the structure. Therefore, the MO concept includes also events with MC-like signatures (as defined by Roulliard, 2011), as well as complex structures.

The origin of CMEs on the Sun is associated with various eruptive and dynamical phenomena in the low corona, such as flares, filament eruptions, magnetic reconfigurations, coronal dimmings, EUV waves, jets, post-eruption arcades, *etc*. (see, *e.g.*, review by Green *et al.* 2018). CMEs that do not exhibit any obvious low-coronal signatures (LCS) are called stealth CMEs (*e.g.*, Robbrecht *et al.*, 2009; Ma *et al.*, 2010; D`Huys *et al.*, 2014). According to Nitta and Mulligan (2017), most stealth-referred CMEs are not strictly stealth, thus the term "stealthy" CMEs would be more appropriate, because the apparent lack of some LCSs can be, at least partly, an instrumental effect (Howard and Harrison, 2013). The most frequent and pronounced LCSs are solar flares and eruptive filaments. The association of CMEs with prominence/filament eruptions has been known since decades (*e.g.*, Munro *et al.*, 1979; Sheely *et al.*, 1980; Schmieder *et al.*, 2013; Webb, 2015, Gopalswamy *et al.*, 2015; Lugaz, 2015). According to the comprehensive statistical study of McCauley *et al.* (2015), filaments can by divided into four types: 1) active region (AR) filaments, *i.e.*, filaments which are found within active regions, 2) intermediate prominences (IP), *i.e.*, filaments located adjacent to or between ARs, 3) quiescent (QS) filaments, *i.e.*, filaments located well away from ARs in quiet sun regions, and 4) polar crown (PC) prominences located at high latitudes.

In this paper, we examine whether and how CMEs associated with different types of filaments produce different *in situ* signatures at Earth. We classify CME eruptions based on their association with filaments, following the filament classification by McCauley *et al.* (2015). We then analyse in detail the corresponding solar wind parameters (flow speed, *v*, magnetic field, *B*, proton density, $N_p$, plasma-*β*, and thermal speed, $V_{th}$) in several regions of the corresponding ICME. The paper is organized as follows: Section 2 describes the data and techniques used to identify isolated Earth-impacting CME/ICME pairs. In Section 3, we analyse CME source locations, classify them based on their association with filaments, and study the particularities of the corresponding ICME *in situ* signatures. In Section 4, we present a detailed analysis of the SW parameters in different regions of the ICMEs, inspect how they differ regarding the types of associated filaments, and discuss the outcome. In Section 5, the results are briefly summarized.

## 2. Data and Method

To identify reliable CME/ICME pairs, we combined CME kinematic measurements with *in situ* data from Wind. Firstly, we visually identified all CMEs using image sequences of the outer coronagraphs COR2 (Howard *et al.*, 2008) onboard STEREO-A and STEREO-B as well as LASCO C3 (Brueckner *et al.*, 1995) onboard SOHO, to estimate whether the CME is oriented towards Earth or not. We selected the time period from January 2008 to August 2014 based on the availability of data from LASCO/C3 and both STEREO-A and STEREO-B COR2 coronagraphs (in October 2014, communication with the STEREO-B spacecraft was completely lost, but already from the very beginning of September 2014 the data required for this study was not useful). We cross-checked all our identifications (in total 1644 CMEs) with that of the SOHO LASCO (https://cdaw.gsfc.nasa.gov/CME_list/) and CACTus (http://sidc.oma.be/cactus/) CME catalogues. In the following we focus only on events that were observed by all three spacecraft.

For each CME selected, we estimated its position angle (PA) from the white light coronagraph observations as the angular position of the point on the CME bright leading edge at the middle of the opening angle, measured counter-clockwise from the north. Using the PAs of CMEs estimated from STEREO-A and -B COR2 images, we selected potentially Earth-directed CMEs, taking into account that when STEREO-A and-B are in quadrature with SOHO, back-sided CMEs have PAs in the range 180° - 360° in the COR2-A FOV, and 0° - 180° in the COR2-B FOV. We applied such a procedure only when the separation angle between STEREO-A and -B was larger than 60°. After isolating the front-side CMEs, we estimated their propagation direction by taking into account the location of the associated low coronal signatures (Section 4).

Figure 1 illustrates the procedure for the CME which erupted on 15 February 2011. In this case, the longitudinal separation between the STEREO-A and STEREO-B spacecraft was ≈181°, whereas the separation angles with respect to the Earth were ≈ 94° and ≈ 87°, respectively. The directions of motion are East, Halo and West in the STEREO-A COR2, LASCO C3 and STEREO-B COR2 coronagraphs, respectively. The source region was located approximately 5° west of the Sun-Earth line. Furthermore, after taking into account the average CME opening angle based on the STEREO-A and -B observations, uncertainties in estimating the source region coordinates, and the possibility of non-radial propagation as well as possible CME deflection, we estimated that if the CME is launched within ±40° from the Sun-Earth line, it is potentially an Earth-impacting CME. In this way, we found that out of the identified 1644 CMEs, 439 (∼27%) were potential candidates for the Earth-impacting ICMEs (Table 1). A list of CMEs observed by all three spacecraft, containing basic information on all 1644 events is published online at https://zvjezdarnica.hr/pdf/ListICMEs.pdf.

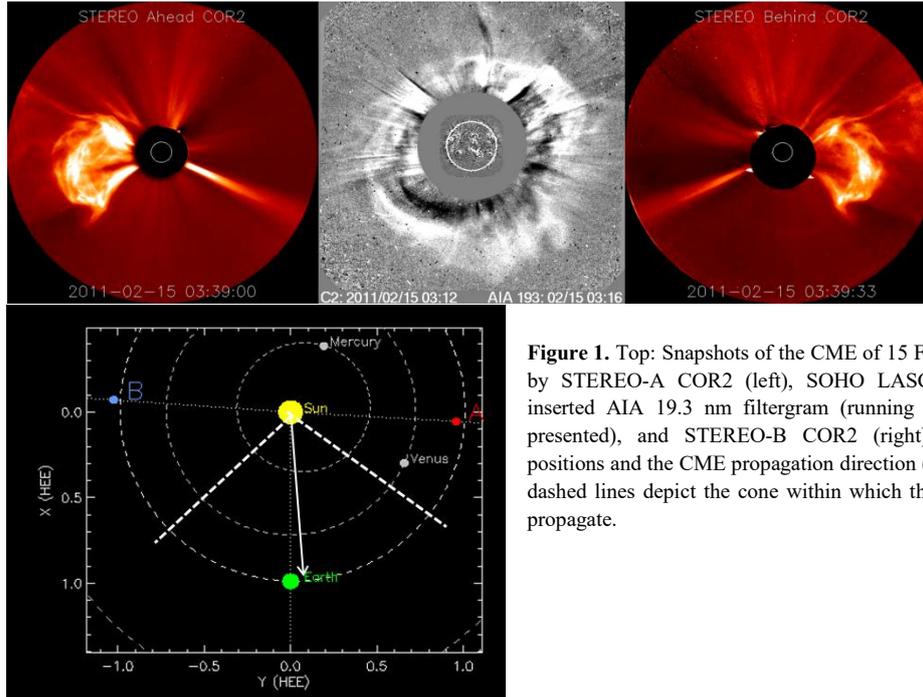

**Figure 1.** Top: Snapshots of the CME of 15 February 2011 provided by STEREO-A COR2 (left), SOHO LASCO C2 (middle) with inserted AIA 19.3 nm filtergram (running difference images are presented), and STEREO-B COR2 (right). Bottom: spacecraft positions and the CME propagation direction (white arrow). The two dashed lines depict the cone within which the CME is expected to propagate.

In the next step, we performed kinematical measurements for every potentially Earth-directed CME in order to estimate the possible arrival time at Earth. For that purpose, we used the white-light images from STEREO-A and -B. The two STEREO spacecraft are equipped with the SECCHI instrument suite (Howard *et al.*, 2008), consisting of an *Extreme Ultraviolet Imager* (EUVI; Wuelser *et al.*, 2004), two white light coronagraphs, the inner COR1 and outer COR2 (Thompson *et al.*, 2003; Howard *et al.*, 2008), and the *Heliospheric Imagers* (HI1 and HI2; Eyles *et al.*, 2009). In our study, we used only data from COR1, COR2 and HI1, since in the EUVI and HI2 FOV many of the studied events were too dim or diffuse for sufficiently accurate kinematical measurements. The kinematical measurements are based on the elongation-time data, $\varepsilon(t)$, of the CME leading-edge in the ecliptic plane, based on visual estimates. The identification of a CME front in images recorded by the instruments listed was accomplished by comparing their morphology in successive snapshots and by comparing and matching their elongation-time curves (illustrated in Figure 2 for the 12 July 2012 CME; for more details of the matching procedure see Maričić *et al.*, 2014). We note that in cases where the CME drives a shock, we used the shock front signature as the leading-edge segment of the CME. For the CME tracking, we used running-difference images instead of J-maps, as they provide a better overview of the morphology and the evolution of the whole CME front during its propagation.

**Table 1** Number of CMEs observed by all three spacecraft and solar wind disturbances recorded by the Wind spacecraft in the period January 2008 to August 2014

| List | Number of CMEs |
| --- | --- |
| CME list | 1644 |
| SWD list | 282 |
| Non Earth-directed CMEs | 1205 |
| Earth-directed CMEs, *i.e.* candidates for Earth-impacting ICMEs | 439 |
| Confirmed Earth-impacting ICMEs | 103 |
| Isolated Earth-impacting ICMEs | 31 |

For the conversion of elongations to radial distances, we applied the standard harmonic mean (HM) fitting method developed by Lugaz (2010) for the stereoscopic observations of ICMEs (see Lugaz, 2010; Section 3). For the application of this method see, *e.g.*, Möstl *et al.* (2011), Temmer *et al.* (2012), Harrison *et* *al.* (2012), Rollett *et al.* (2012), and Maričić *et al.* (2014). In addition, we applied the conversion based on the assumption that the CME is relatively narrow and lies in the plane-of-sky (PoS). Thus, the radial distance is defined as $R = D \tan(\varepsilon)$, where $D$ is the distance from the Sun to the observer (hereinafter, the "PoS method", Maričić *et al.* 2014). Note that the PoS conversion method is analogous to the so-called fixed-$\phi$ conversion (Sheeley *et al.*, 1999; Kahler & Webb, 2007; Rouillard *et al.*, 2008) for a propagation direction of $\phi = 90°$. Then, the linear fits to the HM and PoS data were extrapolated, to calculate the expected arrival time of the ICME at the Wind spacecraft. We emphasize that for the linear-fit extrapolation we used only data from the COR2 and HI1 STEREO instruments. In Figure 2, we illustrate this method by showing elongation-time curves, $\varepsilon(t)$ of the 12 July 2012 CME leading-edge measured in the ecliptic plane (Figure 2g), together with the calculated and extrapolated ICME kinematics (Figure 2h).

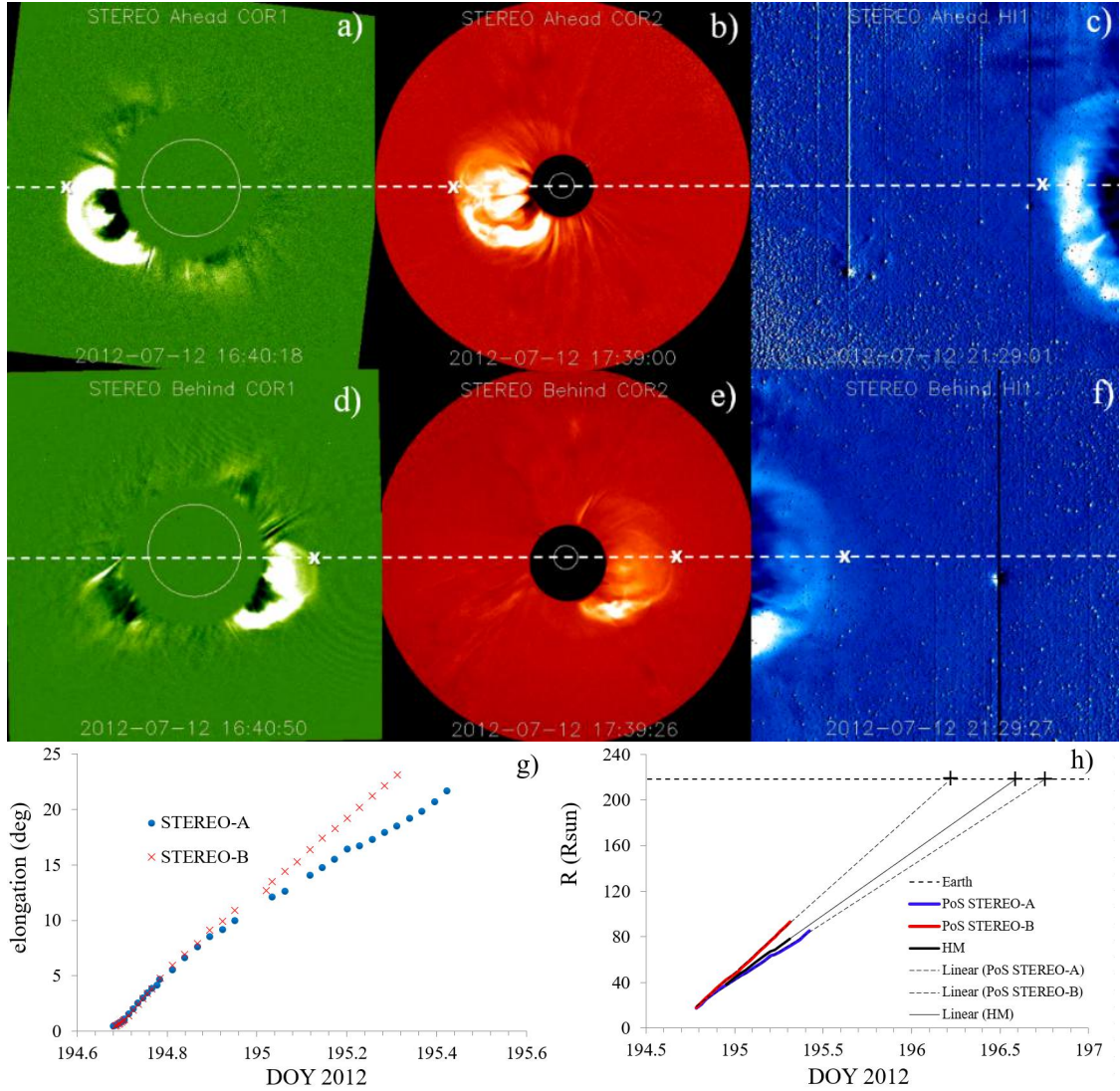

Figure 2. a) - c) STEREO-A COR1 (left), COR2 (middle) and HI1 (right) images of the 12 July 2012 CME. The measured leading-edge segment of the CME is marked by white crosses. The ecliptic plane is indicated by the dashed line. d) - f) Leading-edge identification on STEREO-B images. g) Elongation-time curves, $\varepsilon(t)$, of the leading edge in the ecliptic plane. The STEREO-A and -B observations are marked by blue dots and red crosses, respectively. h) Measured and extrapolated ICME kinematics; the horizontal dashed line represents the heliocentric distance of 1 AU. The blue, black and red bold lines represent the PoS STEREO-A, harmonic mean (HM), and PoS STEREO-B results, the dashed lines represent the linear-fit extrapolation of the PoS STEREO-A and -B data, whereas the solid black line shows the HM extrapolation. The expected ICME arrival at 1 AU, as calculated by the PoS and HM methods, are marked by pluses.

According to previous studies, ICMEs *in situ* signatures can be identified applying various criteria, usually considering smoothness, enhancement, and/or rotation of magnetic field, low plasma-$\beta$, expanding speed profile, presence of a heated and compressed shock/sheath region, *etc.* (see, *e.g.*, Zurbuchen &

Richardson, 2006; Richardson and Cane, 2010; Kilpua *et. al*, 2013; Mitsakou and Moussas, 2014; Paouris and Mavromichalaki, 2017; Kilpua *et al*., 2017 and references therein). In this study, we searched for any solar wind disturbance (SWD) in the *in situ* data within a ±30 hours period around the estimated ICME arrival time, *i.e.*, we looked for deviations from the quiet solar wind conditions in the four basic solar wind parameters (magnetic field strength, flow speed, proton density and proton thermal speed). In order to mark an event as a SWD, we required that it shows a notable enhancement of some of the above mentioned basic solar wind parameters. In addition, we also checked magnetic field components and fluctuations, as well as the plasma-$\beta$ parameter, as indicators of the ICME magnetic structure. We note that the identified SWD signatures are not necessarily characterized by a period of low density, low temperature, low plasma-$\beta$ and smooth rotation of the magnetic field. These signatures may not be present if the spacecraft was not hit by the apex part of the ICME, as the ICME signatures can be different for different ICME trajectories (Gopalswamy, 2006). The SWD with a sheath and/or some other ICME signature closest in time to the estimated arrival of the CME was associated with the CME as its interplanetary counterpart.

To identify ICMEs, we used *in situ* plasma and magnetic field 1-min resolution data from the Wind spacecraft (https://wind.nasa.gov/mfi_swe_plot.php), namely the data from the Solar Wind Experiment (SWE) and the Magnetic Field Investigation (MFI). Data from the ACE satellite (http://www.srl.caltech.edu/ACE/ASC/level2/index.html) was used only for double checking, *i.e.*, only for confirmation of the identified solar wind disturbance in the Wind data. To crosscheck our list of Earth-impacting ICMEs, we consulted several other ICME lists:
- the Wind ICME Catalog
(http://wind.nasa.gov/ICMEindex.php; see Nieves-Chinchilla *et al.*, 2018),
- the University of Science and Technology of China (USTC) List of Interplanetary Coronal Mass Ejections (http://space.ustc.edu.cn/dreams/wind_icmes; see Chi *et al.*, 2016),
- the George Mason University CME/ICME List (http://solar.gmu.edu/heliophysics/index.php/GMU_CME/ICME_List; see Hess and Zhang, 2017),
- the Richardson and Cane Catalog of Near-Earth Interplanetary Coronal Mass Ejections Since January 1996 (http://www.srl.caltech.edu/ACE/ASC/DATA/level3/icmetable2.htm; see Cane and Richardson, 2003 and Richardson and Cane, 2010).

In Figure 3, we illustrate the method how we joined the CME kinematic measurements with the corresponding SWD *in situ* data. Based on the CME trajectories derived from the HM fitting method, we extrapolated the kinematical curves to 1 AU. Out of 439 Earth-directed CMEs, only 103 revealed some of the SWD *in situ* signatures in the expected arrival time period at L1, and only 65 showed ICME signatures (Table 1). The main reason for such a small fraction of CMEs that impacted Earth (about 23%) is most probably a relatively small angular width of the CMEs/ICMEs and/or a propagation out of the ecliptic plane.

Using the derived CME kinematics, we were also able to identify potential CME-CME interactions. The criterion for an CME/ICME to be designated as "interacting" is that its kinematical curve intersects with the extrapolated kinematical curve of another CME/ICME, which should be accompanied by a complex profile in the *in situ* data. In Figure 4, we show an example of interacting CMEs which erupted on 12, 13 and 14 June 2012. For similar studies of CME–CME interactions we refer to the studies by, *e.g.*, Liu *et al.* (2012), Shen *et al.* (2012), and Temmer *et al.* (2012). In the following, we discard all potential interaction events from the sample, and thus establish a final list of 31 isolated Earth-impacting CME/ICME pairs (Table 2) for our study.

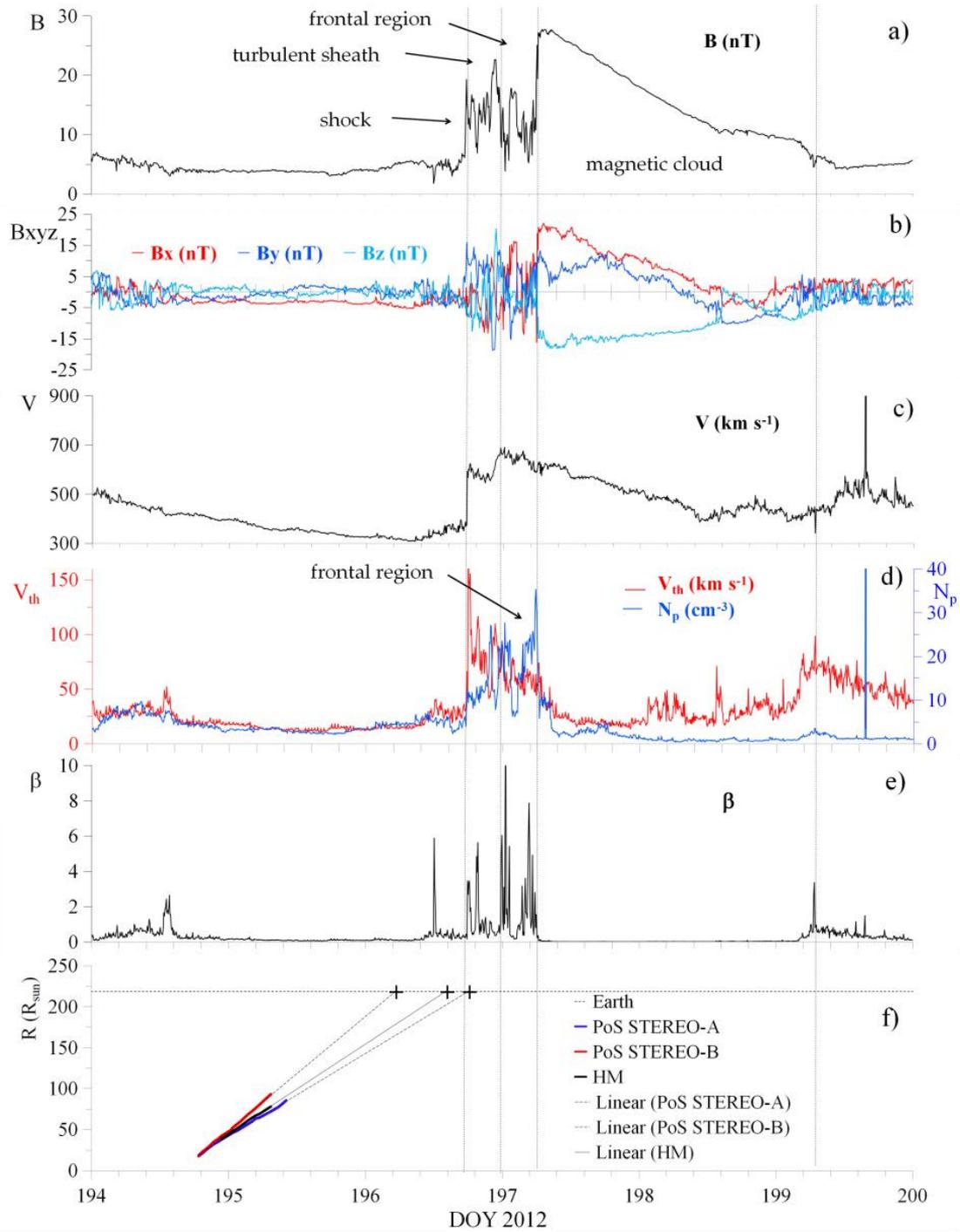

**Figure 3** Example of an isolated Earth-impacting MO-ICME on 13 July 2012 (DOY = 194). a)-e) *In situ* Wind measurements at L1: a) magnetic field magnitude; b) three components of the magnetic field vector in the GSE coordinate system; c) solar wind speed; d) proton density and thermal velocity; e) plasma-$\beta$. Calculated and extrapolated ICME kinematics is shown in panel f. The expected ICME arrival at 1 AU, calculated by the PoS and HM methods, are marked by pluses (as in Figure 2h). Vertical dashed lines in panels a-e represent the borders between different parts of the ICME, *i.e.*, the turbulent sheath, the frontal region, and the magnetic obstacle (for detailed explanation of each region see the main text).

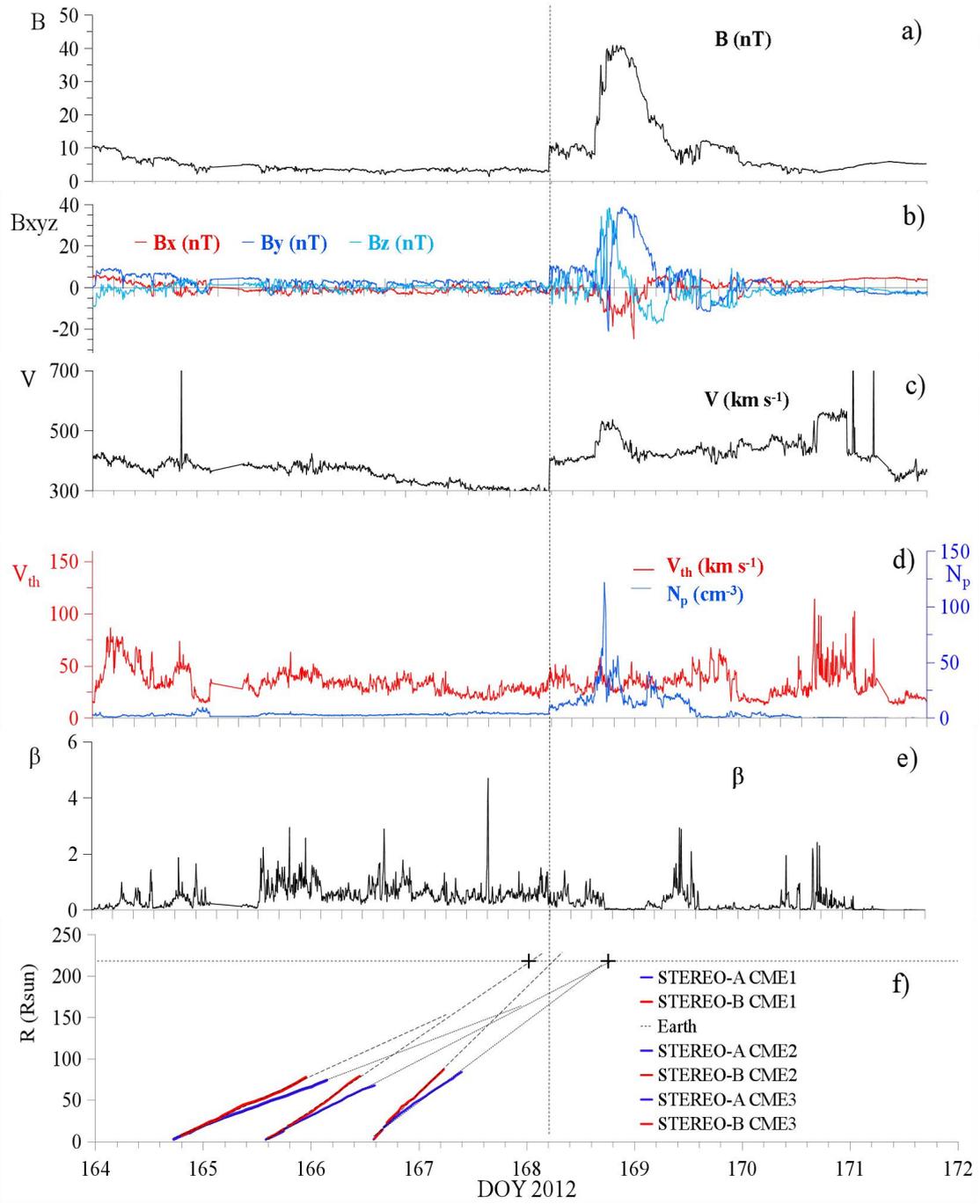

**Figure 4** Example of an ICME interaction event. The CMEs that erupted on 12, 13 and 14 June 2012 (DOY = 164 - 166), are denoted as CME1, CME2, and CME3, respectively. a)-e) *In situ* Wind measurements at L1: a) magnetic field strength; b) GSE magnetic field components; c) solar wind speed, d) proton density and thermal velocity; e) plasma-*β*. Kinematics of the three ICMEs is displayed in panel f, where the horizontal dashed line represents the heliocentric distance of 1 AU. The blue and red bold lines represent PoS STEREO-A and -B CME measurements, whereas the dashed lines represent the corresponding linear-fit extrapolations. The expected arrival of the composite ICME at 1 AU is marked by pluses. The arrival time was determined based on the kinematical curve of the fastest CME in the group.

## 3. Source Region Locations and Association with Filament Eruptions

For the 31 isolated Earth-impacting ICMEs that were identified in the period 2008 – 2014 (Section 2), first we analyse their low coronal signatures in more detail. The main reason for this additional analysis is to identify the solar source location, the association with filament eruptions and flares, and to analyse the CME propagation direction.

To this aim, STEREO coronagraphic data are combined with SDO extreme-ultraviolet and soft X-ray (SXR) data to identify potentially associated solar flares, since the acceleration phase of a CME is often closely synchronized with the energy release in the associated flare (Zhang *et al.*, 2001; Vršnak *et al.*, 2004, Maričić *et al.*, 2007; Temmer *et al.* 2008; 2010, Berkebile-Stoiser *et al.,* 2012, Veronig *et al.*, 2018). The heliographic location of flares (used in the subsequent analysis as the source location of the CMEs) was determined using the *Extreme ultraviolet Variability Experiment* (EVE; Woods *et al.*, 2012) onboard SDO, *i.e.*, the SAM pinhole camera images and the SXR flux in the 0.1–7 nm channel (http://lasp.colorado.edu/eve/data_access/). Two CMEs (9 July 2013 and 14 June 2011) were not associated with any notable SXR flare. For these two events, the CME solar source location was determined using SDO/AIA data, by measuring the location of the central part of the associated eruptive filament. CMEs which are not associated with a prominence eruption, flare emission enhancement, or a post-flare arcade, are referred as stealthy CMEs (Nitta and Mulligan, 2017). Out of the final set of 31 isolated Earth-impacting ICMEs, four turned out to be stealthy CMEs. Consequently, in the case of these four CMEs we were not able to pin-point their origin on the Sun.

We divided the isolated Earth-impacting ICME sample into different types with respect to the type of the associated eruptive filament:

1) Active region (AR) CMEs: These are CMEs that have their origin inside an AR and that are associated with an AR filament type. In our data set, all CMEs of this type are associated with GOES X-class flares.
2) Disappearing filament (DSF) CMEs: These are CMEs whose origin is associated with disappearing filaments of QS or IP type. These filaments can be connected to ARs, but they are not located inside the AR. The most powerful SXR flare related to this CME type was of GOES class M1.3.
3) Stealthy CMEs: These are CMEs which cannot be related to any filament activation, nor to any significant increase (stronger than GOES A-class flare) of the SXR flux.

The type of event regarding this classification is presented in the last column of Table 2.

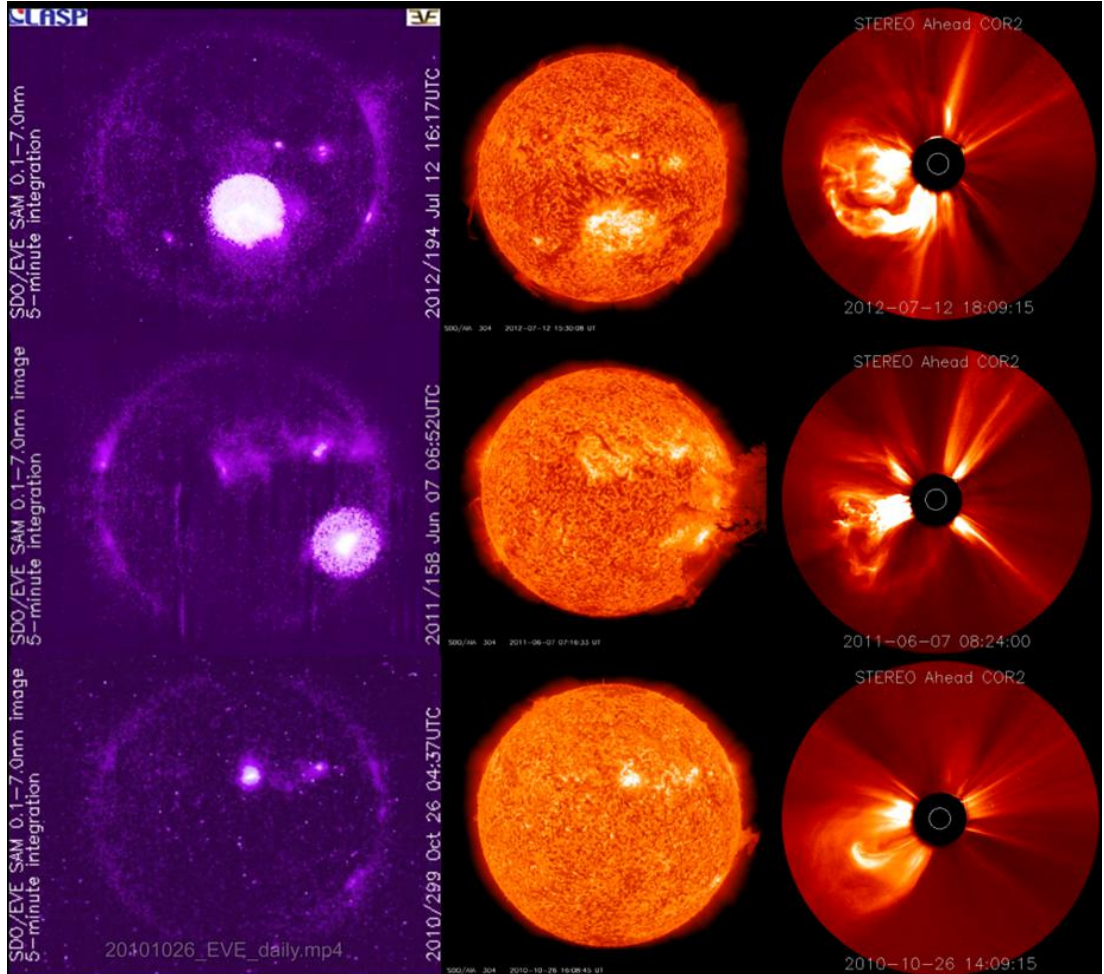

**Figure 5** Three examples of isolated Earth-impacting ICMEs (from top to bottom: 12 July 2012, 7 June 2011, and 26 October 2010), recorded by the SDO/EVE SAM pinhole camera, the SDO/AIA 304 A filter, and the COR2 STEREO-A coronagraphs (from left to right). The 12 July 2012 event represents an eruption with its origin close to the central meridian. The 7 June 2011 event has its origin close to the solar limb. The 26 October 2010 eruption is a "stealthy" CME, showing no low-coronal signature.

According to Vourlidas *et al.* (2013), the "classical three part CME" morphology visible in white-light coronagraph images (Illing and Hundhausen, 1985) can be updated to a "five-part CME" structure. In the traditional view, the CME consists of a dark cavity, which is surrounded by a bright frontal loop, and an embedded bright core that is generally assumed to be an erupting prominence (however, for different interpretations see Howard *et al.*, 2017 and Veronig *et al.*, 2018). Vourlidas *et al.* (2013) added a two-front morphology, where the outer bright loop is preceded by a faint sharp front and a broad region of diffuse emission. These features were interpreted as a proxy of the CME-driven shock wave and the shock-associated sheath compression. We adopt a similar reasoning for the *in situ* ICME measurements. Furthermore, we noted that in the ICME structure attached to the frontal edge of the magnetic obstacle (MO), *i.e.*, at the end of the shock-associated turbulent sheath (TS), a region characterised by an additional enhancement of proton density and flow speed can be often recognized. This enhancement is denoted in Figures 3a and d as "frontal region" (hereafter, FR). In the following analysis, this part of the ICME is treated as a separate segment of the ICME.

Next, we analyse whether or not the *in situ* signatures in some part of the ICME show magnetic cloud (MC) properties, *i.e.*, contain a region of a smooth rotation of the magnetic field direction, low proton temperature, and low proton plasma-$\beta$ (Burlaga *et al.*, 1981; Burlaga, 1988, 1995; Burlaga *et al.*, 2008, Kim *et al.*, 2013, Nieves-Chinchilla *et al.* 2018, 2019). In this respect, we adopt the terminology used by Jian *et al*. (2006) and Nieves-Chinchilla *et al*. (2018), where an ICME is denoted as a "magnetic obstacle" (MO) if even a small region with typical MC signatures is identified, *i.e.*, when the ICME configuration is not associated with a full single magnetic field rotation. Following this terminology, we divided the sample of Earth-impacting ICMEs into two categories:

1) We denote an ICME as a "magnetic obstacle" (MO) if it contains a structure with MC or MC-like properties as defined by Rouillard (2011), *i.e.*, corresponding to ICMEs of Group 1 according to Jian *et al.* (2006) and F-type according to Nieves-Chinchilla *et al.* (2019).

2) An ICME is termed as "ejecta" (EJ) if it does not show any MC or MC-like properties, *i.e.*, corresponds to Groups 2 and 3 according to Jian *et al.* (2006), E-type according to Nieves-Chinchilla *et al.* (2019), as well as events with typical shock/sheath signatures but without MO signatures.

An example of a MO-ICME is given in Figure 3 and of an EJ-ICME in Figure 6. The MO/EJ types are listed in Column 16 of Table 2.

We consulted also ICME lists by Richardson & Cane (2010), the Wind ICME catalogue, the HELCATS LINKCAT catalogue (https://www.helcats-fp7.eu/catalogues/wp4_cat.html), and the ICME list by the University of Science and Technology of China, and we compared them with our list. We note that some inconsistencies among the mentioned three lists were found. After a more detailed inspection we inferred that the detected inconsistencies are either due to differences in terminology, or sometimes, they appeared in the cases of ICMEs without clear MC signatures where it was not taken into account that the event was characterized by a flank encounter.

The main characteristics of the identified isolated Earth-impacting ICMEs are listed in Table 2. Column 1 gives the event label, in Columns 2 and 3 the times of the first CME detection in STEREO COR1 are displayed, Columns 4 - 9 present the onset, magnetic field and plasma parameters of the related *in situ* SWD, Columns 10 - 12 define the location and GOES SXR importance of the associated flare, Column 13 specifies the associated active region, Columns 14 and 15 provide information on the filament type and angular width. The filament type is acquired from the AIA filament eruption catalog (http://aia.cfa.harvard.edu/filament/), where the classification by McCauley *et al.* (2015) is used: active region (AR) filaments, intermediate prominences (IP), *i.e.*, filaments located adjacent to or between ARs, quiescent (QS) filaments, and polar crown (PC) filaments. We note that two events (2011 Mar 25 and 2012 May 11) were not found in the AIA filament eruption catalog, thus we inspected AIA images and Kanzelhohe Observatory filtergrams (http://cesar.kso.ac.at/halpha3s/). In this way, we determined that these two events were related to IP filaments. Let us note that no event was associated with a PC filament. Finally, column 16 lists the MO/EJ ICME type, and column 17 the CME source-region type. In column 16, the MOs are identified following the criteria of Nieves-Chinchilla *et al.* (2018, 2019).

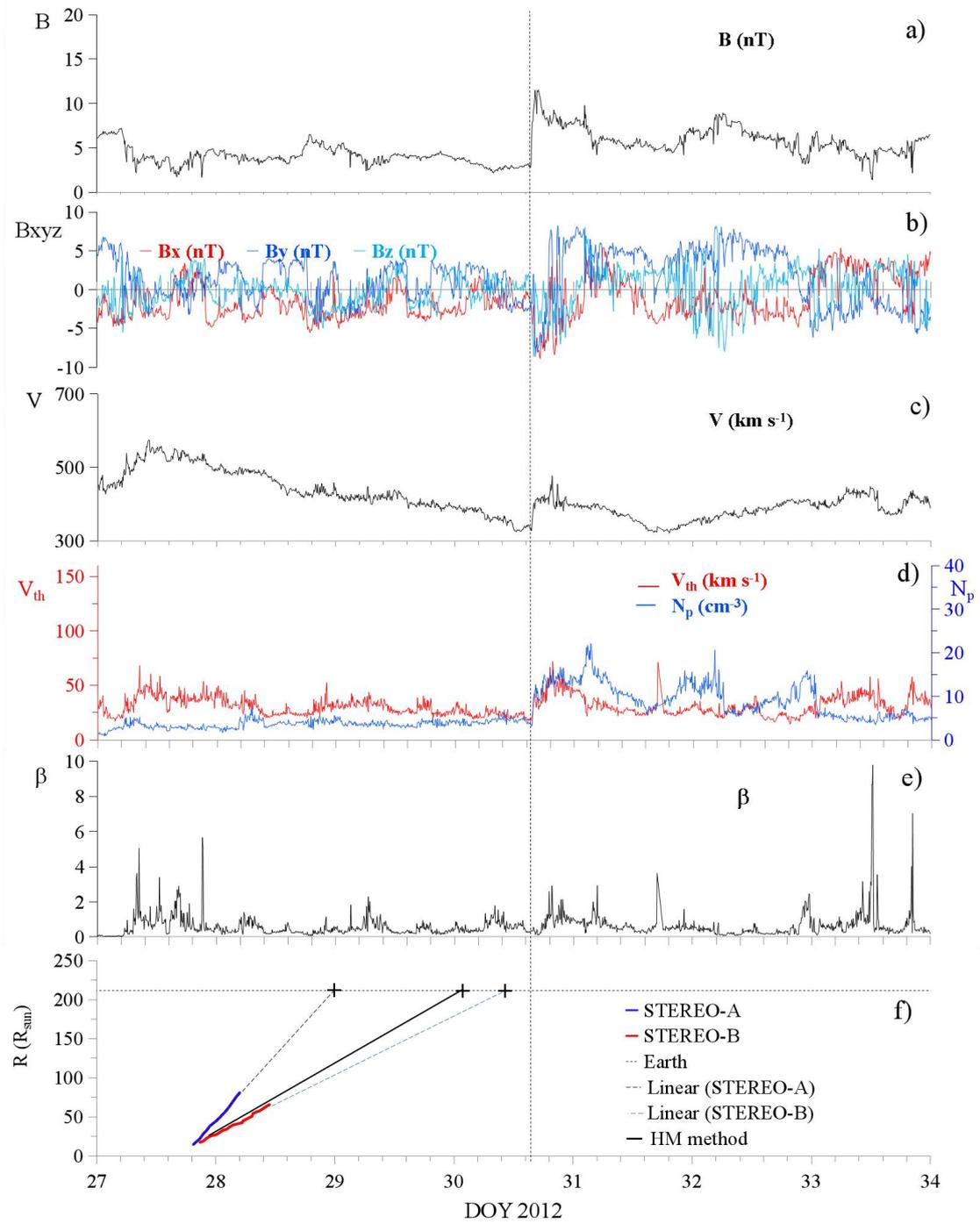

**Figure 6** Example of an EJ-ICME showing *in situ* Wind signatures at L1 (DOY = 27 is 27 January 2012). a) magnetic field strength; b) GSE magnetic field components; c) solar wind speed; d) proton density and thermal velocity; e) plasma-$\beta$; f) CME/ICME kinematics. The horizontal dashed line in panel f represents a heliocentric distance of 1 AU. The blue and red bold lines represent PoS STEREO-A and -B CME measurements, the dashed lines represent the linear-fit extrapolations to 1 AU, whereas the solid black line shows the result of the harmonic mean (HM) fitting method. Vertical dashed line marks the onset of the ICME-driven shock. The predicted arrival times of the ICME at 1 AU calculated by the PoS and HM methods are marked by pluses.

Table 2 Parameters of the isolated Earth-impacting ICMEs during the period from January 2008 to August 2014

| No. | Earliest recording of the CME STEREO A COR1 (UT) start date and time | | ICME onset at Wind (UT) start date and time | | B (nT) | V (kms$^{-1}$) | Np (cm$^{-3}$) | vth (kms$^{-1}$) | Flare location | | Flare SXR type | AR | Filament type | Filament angular width (°) | MO | CME source type |
|---|---|---|---|---|---|---|---|---|---|---|---|---|---|---|---|---|
| 1  | 2010 Apr 3  | 9:15  | 2010 Apr 5  | 11:46 | 19.24 | 815.6 | 16.2 | 147.7 | S23 | E03 | B7.4 | 11059 | AR | 0.0107 | MO | AR |
| 2  | 2010 Aug 7  | 18:20 | 2010 Aug 10 | 14:09 | 6.2   | 478.4 | 12.1 | 64.2  | N11 | E34 | M1.0 | 11093 | AR | 0.0234 | EJ | AR |
| 3  | 2010 Oct 26 | 3:50  | 2010 Oct 30 | 9:08  | 11.54 | 395.4 | 23.2 | 44.2  | ×   | ×   | ×    | ×     | ×  | ×      | MO | stealthy |
| 4  | 2011 Mar 25 | 6:55  | 2011 Mar 29 | 15:02 | 8.87  | 394.1 | 38   | 92    | S10 | E25 | M7.1 | 11176 | IP | 0.0105 | MO | DSF |
| 5  | 2011 Jun 14 | 5:55  | 2011 Jun 17 | 1:33  | 9.45  | 559.1 | 6.9  | 83.4  | ×   | ×   | ×    | ×     | QS | 0.1620 | EJ | DSF |
| 6  | 2011 Jun 21 | 2:20  | 2011 Jun 22 | 18:53 | 11.1  | 685.2 | 7.7  | 90.4  | N17 | W10 | C7.7 | 11263 | AR | 0.0864 | EJ | AR |
| 7  | 2011 Sep 13 | 21:35 | 2011 Sep 17 | 2:55  | 13.72 | 554.9 | 59.7 | 61.9  | ×   | ×   | ×    | ×     | ×  | ×      | MO | stealthy |
| 8  | 2011 Nov 9  | 13:25 | 2011 Nov 12 | 4:54  | 8.05  | 473.7 | 8    | 63.6  | N24 | E35 | M1.1 | 11343 | IP | 0.0525 | EJ | DSF |
| 9  | 2011 Nov 26 | 6:30  | 2011 Nov 28 | 20:44 | 17.52 | 535   | 22.6 | 110.1 | N11 | W47 | C1.2 | 11353 | IP | 0.1260 | EJ | DSF |
| 10 | 2012 Jan 16 | 3:00  | 2012 Jan 20 | 3:26  | 7.23  | 384.5 | 9.8  | 32.8  | N15 | E88 | C2.1 | 11401 | AR | 0.0420 | EJ | AR |
| 11 | 2012 Jan 27 | 18:20 | 2012 Jan 30 | 15:23 | 11.51 | 417.4 | 12   | 71.9  | N33 | W85 | X1.7 | 11402 | AR | 0.0483 | EJ | AR |
| 12 | 2012 Mar 13 | 17:25 | 2012 Mar 15 | 12:33 | 16.03 | 807.7 | 8.2  | 118.1 | N17 | W66 | M7.9 | 11429 | AR | 0.0315 | MO | AR |
| 13 | 2012 Apr 2  | 0:05  | 2012 Apr 5  | 7:30  | 9.4   | 371   | 23.1 | 96.92 | ×   | ×   | ×    | ×     | ×  | ×      | MO | stealthy |
| 14 | 2012 May 11 | 23:30 | 2012 May 17 | 6:17  | 13.15 | 410.7 | 12.1 | 78    | S15 | E20 | C3.2 | 11477 | IP | 0.0597 | MO | DSF |
| 15 | 2012 May 17 | 1:40  | 2012 May 20 | 1:10  | 7.88  | 465.8 | 8.7  | 47.5  | N07 | W88 | M5.1 | 11476 | AR | 0.0427 | EJ | AR |
| 16 | 2012 Jul 12 | 16:15 | 2012 Jul 14 | 16:11 | 27.2  | 692.6 | 26.7 | 148.9 | S13 | W03 | X1.4 | 11520 | AR | 0.0215 | MO | AR |
| 17 | 2012 Jul 17 | 12:40 | 2012 Jul 20 | 3:52  | 13.4  | 472.3 | 11.9 | 119.4 | S18 | W81 | C2.2 | 11520 | AR | 0.0871 | EJ | AR |
| 18 | 2012 Aug 4  | 13:50 | 2012 Aug 7  | 20:02 | 9.57  | 431.1 | 17.7 | 83.1  | S19 | E55 | C3.5 | 11539 | IP | 0.1188 | EJ | DSF |
| 19 | 2012 Aug 31 | 19:50 | 2012 Sep 3  | 10:55 | 26.38 | 566   | 37.8 | 76.4  | S19 | E45 | M1.3 | 11562 | IP | 0.1259 | EJ | AR |
| 20 | 2012 Oct 27 | 8:25  | 2012 Oct 31 | 14:08 | 15.78 | 347.8 | 38.3 | 34    | S15 | W11 | C1.1 | 11598 | AR | 0.0190 | MO | AR |
| 21 | 2013 Feb 12 | 22:20 | 2013 Feb 16 | 11:15 | 12.17 | 384.8 | 14.2 | 54.3  | S30 | W60 | B2.1 | ×     | QS | 0.0924 | EJ | DSF |
| 22 | 2013 Mar 15 | 6:30  | 2013 Mar 17 | 5:08  | 19.62 | 685.6 | 11.6 | 143.2 | N09 | E06 | M1.1 | 11629 | AR | 0.0614 | MO | AR |
| 23 | 2013 Apr 11 | 7:15  | 2013 Apr 13 | 21:59 | 13.16 | 515.2 | 21.3 | 64.2  | N07 | E13 | M6.5 | 11719 | AR | 0.0448 | MO | AR |
| 24 | 2013 May 15 | 1:55  | 2013 May 18 | 0:13  | 13.88 | 437.9 | 26.8 | 59.2  | N10 | E68 | X1.2 | 11748 | AR | 0.0281 | EJ | AR |
| 25 | 2013 May 17 | 9:10  | 2013 May 19 | 22:20 | 13.62 | 438.1 | 17.9 | 76.4  | N11 | E56 | M3.2 | 11748 | AR | 0.0367 | EJ | AR |
| 26 | 2013 Jun 21 | 3:15  | 2013 Jun 23 | 3:42  | 8.1   | 687.6 | 4.2  | 116   | S14 | E73 | M2.9 | 11777 | AR | 0.0259 | EJ | AR |
| 27 | 2013 Jun 23 | 22:05 | 2013 Jun 27 | 13:32 | 11.67 | 460   | 33.2 | 63.5  | ×   | ×   | ×    | ×     | ×  | ×      | MO | stealthy |
| 28 | 2013 Jul 9  | 14:45 | 2013 Jul 12 | 19:29 | 16.38 | 528.2 | 18.2 | 96.1  | ×   | ×   | ×    | ×     | QS | 0.1317 | MO | DSF |
| 29 | 2013 Aug 17 | 18:50 | 2013 Aug 20 | 21:03 | 11.5  | 575.6 | 24.4 | 62.1  | S07 | W40 | M3.3 | 11818 | AR | 0.0583 | EJ | AR |
| 30 | 2013 Sep 29 | 21:40 | 2013 Oct 2  | 1:02  | 30.12 | 627.1 | 57.4 | 141.3 | N20 | W22 | C1.2 | 11850 | IP | 0.1387 | MO | DSF |
| 31 | 2013 Oct 25 | 15:10 | 2013 Oct 29 | 9:24  | 14.04 | 438.5 | 20.8 | 95.2  | S06 | E69 | X2.1 | 11882 | AR | 0.0339 | EJ | AR |

## 4. Results and related Discussions

Out of the 31 isolated Earth-impacting ICMEs considered in this study, 17 events (~55%) are not associated with a MO, whereas 14 events (~45%) contain a MO structure. Out of the 14 MO events, four were stealthy CMEs, so their source-region location is unknown. We found a significant difference between the source location of MO- and EJ-ICMEs (Figure 7). 90% of the 10 non-stealthy MO-CMEs (9/10) have their source location within ±30° from the central meridian of the Sun; the only exception is the 13 March 2012 event as its source location is N17W66. On the other hand, 94% of EJ-CMEs (16/17) have their source region outside ±30° from the central meridian; the only exception is the 21 June 2011 event with a source location N17W10. Figure 7 shows the distribution of the source region locations of 27 CMEs (four stealthy CMEs are excluded). The mean longitude ($|l|$) of the 10 non-stealthy MO events is East 3.7 ± 2°, whereas for the East and the West population of EJ events the mean central meridian distances are East 58.1° ± 2° and West 69.7° ± 2°, respectively. This clearly shows that the source regions of MO-ICMEs are grouped close to the solar disc centre, whereas EJ-CMEs are grouped closer to the solar limb, indicating that MO-associated CMEs propagate more directly toward the Earth. The existence of events, which have their origin located more than ±30° away from the central meridian but are associated with a MO, or events inside the ±30° region that are not associated with a MO, could be caused by a non-radial CME propagation from the source region, or deflection from their original direction (see, *e.g.*, Möstl *et al.*, (2015) and Kim *et al.* (2013)). These results support the hypothesis that ICMEs show different *in situ* signatures depending on the spacecraft trajectory across the ICME (see e.g. Jian *et al.*, 2006; Richardson & Cane, 2010; Kilpua *et al.*, 2017 and references therein). Finally, we note that the difference between the mentioned mean source region longitudes for the EJ E/W events (~58°E *vs.* ~70°W) is consistent with the so-called eastward deflection, as a significant fraction of the EJ events is faster than the nominals slow solar wind (see Table 2; average speed is 496 ± 22 kms$^{-1}$). Consequently, to hit Earth the East-hemisphere events must have their origin closer to the central meridian than those launched from the West-hemisphere.

EJ-ICMEs with a source location longitude larger than ±30° from the central meridian, cause a specific signature in the solar wind *in situ* data. Figure 6 shows an example of such an ICME signature. The data reveal a simple structure of the interplanetary disturbance, characterized by a simultaneous increase of the magnetic field strength, thermal velocity, proton density, and solar wind speed, and not exposing any clear signature of the rotation of the magnetic field, or a low plasma-$β$. After the initial sharp increase, the flow speed, thermal velocity and magnetic field strength are characterized by a gradual decrease. The average of the peak values of the magnetic field strength, solar wind speed, proton density, and thermal velocity, for the 17 EJ-ICMEs are 11.85 ± 1.61 nT, 496 ± 22 kms$^{-1}$, 15.52 ± 2.11 cm$^{-3}$, and 76.82 ± 5.81 kms$^{-1}$, respectively. Bearing in mind the close-to-limb source-region locations of EJ-ICMEs, one can conclude that they are observed when the ICME passes over the spacecraft by its flank. For comparison, the average of the peak values of the magnetic field strength, solar wind speed, proton density and thermal velocity for the 10 non-stealthy MO-ICMEs are 17.95 ± 2.05 nT, 582 ± 53 kms$^{-1}$, 24.8 ± 4.9 cm$^{-3}$ and 163 ± 13 kms$^{-1}$, respectively, *i.e.*, are significantly higher than in the case of EJ-ICMEs.

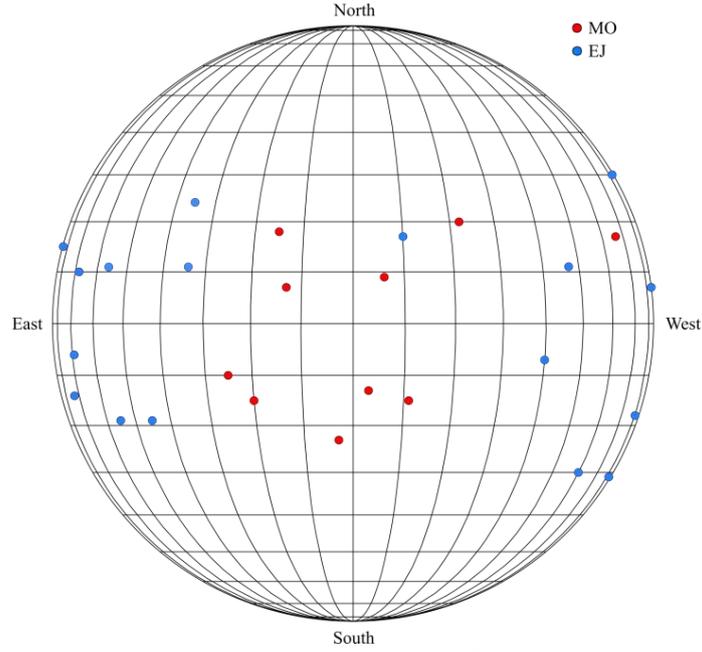

**Figure 7** Distribution of the source region locations for MO-ICMEs (red dots) and EJ-ICMEs (blue dots).

In the next step, a more detailed analysis of the *in situ* Wind measurements of MO-ICMEs was performed. We compared the characteristics of the three structural elements of these ICMEs (cf. Figure 3): the turbulent sheath (TS), the frontal region (FR), and the magnetic obstacle (MO). The second vertical dashed line in Figure 3a-e represents the border between the TS and FR, while the third one represents the end of the FR, *i.e.*, the onset of the MO. In all three ICME segments, we read out the maximum value of the magnetic field strength, the change in the GSE magnetic field components, the flow speed, maximum values of the flow speed, temperature, proton density, and plasma-*β*. Figure 8 shows various correlations between the maximum values of the magnetic field strength, $B_{max}$, in the three MO-ICME regions (TS, FR and MO). In each panel we display the linear least-squares fit and the Pearson correlation coefficient. The highest correlation coefficient *cc* = 0.96 is found for the correlation between the maximum magnetic field strengths in TS and FR, whereas analogous TS/MO and FR/MO correlations are characterised by somewhat lower values, *cc* = 0.87 and 0.86, respectively (Figures 8a-c).

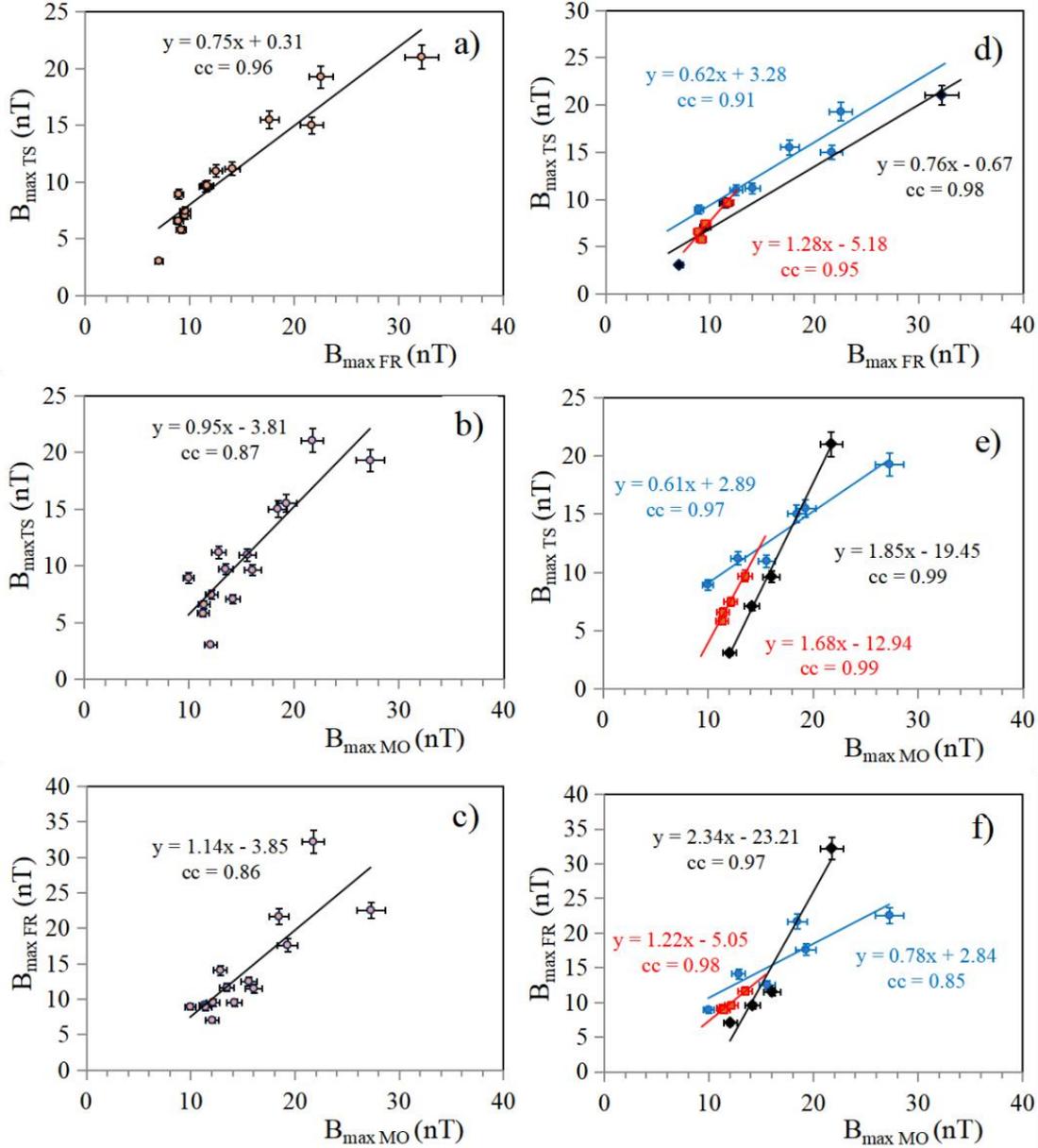

**Figure 8** Correlations concerning the maximum magnetic field magnitude, $B_{max}$, in the three different segments of MO-ICMEs: turbulent sheath (TS), frontal region (FR) and magnetic obstacle (MO). a) TS *vs.* FR; b) TS *vs.* MO; c) FR *vs.* MO. In panels d, e, and f, plots analogous to those in panels a, b, and c are shown, but the ICMEs are divided into different types: AR-events (blue), DSF-events (black), and stealthy CMEs (red). Note that some data-points overlap.

The strong correlation of $B_{max}$ in TSs and FRs is not surprising since these two regions both contain ambient solar wind plasma compressed by the MO motion. The TS represents the downstream region of the shock structure that most often has partly characteristics of a freely propagating wave, and partly is still affected by the MO acting as a piston (for the characteristics and evolution of fast-mode MHD waves see, *e.g.*, Vršnak & Lulić, 2000, and Sect. 2 in Warmuth, 2015). Note that although the characteristics of the shock downstream regions are determined by the Rankine-Hugoniot jump relations, *i.e.*, by the shock amplitude, they are partly affected by the characteristics of the FR, since the plasma flow in the FR still feeds the shock and thus affects the shock amplitude. This results in a strong correlation between TS and FR. On the other hand, although the characteristics of the FR, and thus also the TS are governed by the MO motion, the correlation of TS and FR with MO is somewhat weaker, since the MO structure is a magnetically isolated entity that contains plasma carried from below, whereas the TS and FR segments contain compressed solar wind plasma and "open" magnetic field.

The relations between the different ICME segments were further analysed, focusing on the filament type associated with the CME. According to Table 2, 87% of the isolated Earth-impacting ICMEs in our sample are associated with filament eruptions, whereas 13% (4/31) are not (stealthy CMEs). 55% (17/31) of CMEs are associated with the AR filaments, and 32% (10/31) are associated with QS or IP filament type (*i.e.*, DSF type). In Figures 8d-f, we divided the ICME data set into three different subsets: AR, DSF, and stealthy type. Again, distinct correlations are found between the values of $B_{max}$ for all filament types.

In the case of the TS/FR correlation (Figure 8d), the correlation coefficients for AR, DSF, and stealthy ICME types are similar (*cc* = 0.91, 0.98, and 0.95, respectively). We note that all three types show very similar correlation patterns, *i.e.*, all data-points are grouped in a compact cloud. Such an outcome is not surprising, since both the TS and the FR region consist of compressed ambient solar wind plasma and directly affect each other.

On the other hand, when the TS/MO and FR/MO correlations are considered (Figures 8e and f), one finds a somewhat different situation. Here, the three ICME types reveal significantly different regression lines. In the case of the TS/MO correlation, the DSF-sample shows a twice steeper regression line than the AR-sample, whereas the stealthy-sample behaves similarly to the AR-sample. All three correlations (DSF, AR, stealthy; Figure 8e) have high correlation coefficients, *cc* = 0.97, 0.99, and 0.99, respectively. In the case of the FR/MO correlation, shown on Figure 8f, one finds a similar situation: the DSF-sample again shows a twice steeper regression line than the AR-sample. The correlations coefficients are *cc* = 0.98, 0.85, and 0.97 for the DSF, AR, and stealthy ICMEs, respectively.

Different linear dependencies of $B_{max}$ found for AR and DSF types in the TS/MO and FR/MO correlations suggest that these three ICME types differ probably due to the fact that they originate from different solar magnetic field environments. Events coming from ARs should carry stronger magnetic field than those whose origin is in quiet sun regions. Thus, MO of AR-ICMEs tend to have stronger fields than DSF-ICMEs (see Table 3 and Figures 8e and f), whereas the ambient solar wind field, composing the TS and FR segments, should be similar in both ICME types. Consequently, the considered magnetic field ratio is different, leading to a different inclination of the regression lines. Following this logic, the similarity of the correlations for DSF and stealthy ICME types shown in Figure 8e suggest that stealthy ICMEs are probably launched from quiet sun regions. Note also that the difference between the correlations in both Figure 8e and Figure 8f also enlarged the scatter of data points, *i.e.*, decreased the correlation coefficients in Figures 8b and c.

In Figure 9, we present the TS/FR, TS/MO, and FR/MO correlations, regarding the bulk solar wind speed, $v_{max}$, the proton density, $Np_{max}$, and the proton thermal speed, $v_{thmax}$. The strongest correlations are found for $v_{max}$, characterised by *cc* = 0.96, 0.93, and 0.94, for TS-FR, TS-MO, and FR-MO correlations, respectively (Figures 9a-c). Such an outcome is expected since the flow speeds in TS and FR are physically tightly connected. The FR compression is directly attached to the MO, thus both must have the same speed at the contact surface. This results in the linear regression slope, *k*, close to *k* = 1 (more specifically, *k* = 0.91, see Figure 9c). Similarly, the shock amplitude, thus also the flow speed in the shock downstream region are governed by the MO speed. However, in the case of the bow-shock type, i.e., the shock at least partly driven by the MO, this speed is expected to be lower than that of the MO itself, which resulted in *k* < 1 for the TS/MO correlation and consequently, also for the TS/FR correlation (more specifically, *k* = 0.86 and 0.82 for TS/FR and TS/MO correlations, respectively, see Figures 9a and b).

For $Np_{max}$, a significant correlation is found only for the TS/FR correlation, with *cc* = 0.78 (Figure 9d), which is a consequence of a tight physical connection between these two regions. Similarly, in the case of the thermal speed, $v_{thmax}$, a strong correlation is found for the TS/FR correlation (*cc* = 0.95, see Figure 9g). The TS/MO and FR/MO relations show a significantly weaker correlation coefficients (*cc* = 0.77 and 0.84, respectively, see Figures 9h and i) as expected, as TS and FR segments are representing magneto-plasma systems different from that of MO.

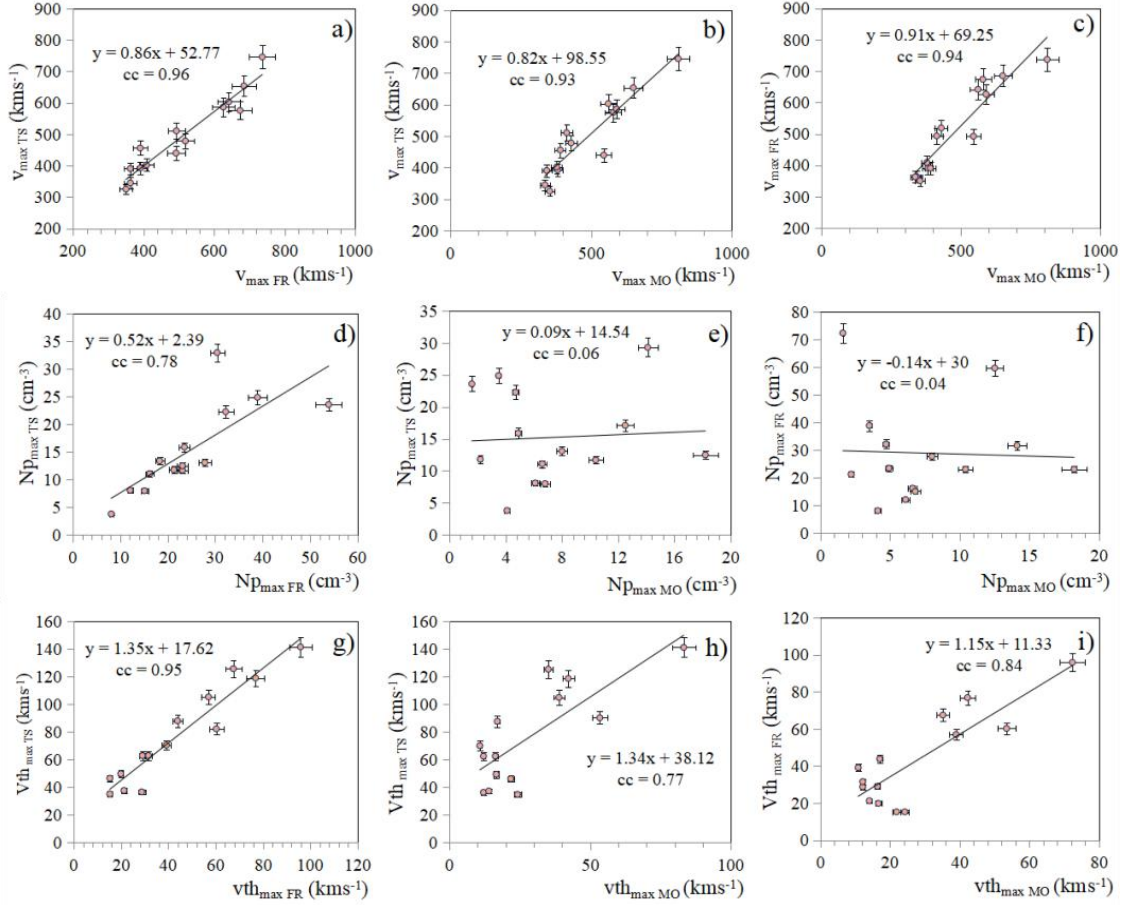

**Figure 9** TS *vs*. FR, TS *vs*. MO, and FR *vs*. MO correlations of: a) - c) the solar wind speed, $v$; d) - f) the proton density, $Np$; g) - i) the proton thermal speed $v_{th}$.

**Table 3**

Mean values of the maximum magnetic field strengths, flow speeds, proton densities and thermal speeds for three different MO-CME types. Values are estimated using the Wind *in situ* data.

| Type of MO-CME | Number of Events | Average values of | | | |
|---|---|---|---|---|---|
| | | $B_{max}$ (nT) | $v_{max}$ (kms$^{-1}$) | $Np_{max}$ (cm$^{-3}$) | $v_{thmax}$ (kms$^{-1}$) |
| *AR* | 6 | 18.50 ± 1.99 | 645 ± 74 | 20.38 ± 4.49 | 109.35 ± 19.97 |
| *DSF* | 4 | 17.13 ± 4.59 | 490 ± 44 | 31.42 ± 10.27 | 101.85 ± 10.71 |
| *Stealthy* | 4 | 11.58 ± 0.88 | 445 ± 41 | 34.8 ± 8.6 | 67 ± 11 |

## 5. Summary

In this paper, we analysed the characteristics of isolated Earth-impacting ICMEs by combining remote sensing observations of CMEs and associated flares and filaments using the STEREO-SECCHI, SOHO-LASCO, SDO-EVE and SDO-AIA instruments, in combination with the *in situ* Wind plasma and magnetic field measurements. We identified 31 isolated Earth-impacting ICMEs, which we divided into those containing a Magnetic Obstacle (MO) structure in the *in situ* data, and those that do not show an MO signature (ejecta, EJ). A detailed analysis of the magnetic field and plasma characteristics in different MO-ICME segments was performed. In addition, we classified the ICMEs into different types according to their solar source region and filament association, distinguishing active region CMEs (AR), quiet sun CMEs (DSF), and stealthy CMEs. The main results and interpretations of our study are as follows.

a) A significant difference was found between the solar source location of MO-ICMEs and EJ-ICMEs. In the vast majority of cases, MO-ICMEs source regions are grouped close to the solar disc centre (90% of them are located at a central meridian distances (CMD) within ±30°), indicating that they are directed more or less towards the Earth. On the other hand, the majority of EJ-ICMEs originate from regions located closer to the solar limb (94% of them are found out of a CMD distance of ±30°, with a mean CMD of 65°).

b) Whereas *in situ* signatures of MO-ICMEs are characterized by the presence of a flux rope in at least some part of their structure, the EJ-ICMEs do not contain such a signature. This indicates, together with item a), that EJ-ICMEs hit the Earth with the ICME flank, whereas MO-ICMEs hit the Earth with the ICME apex, which is also consistent with item a).

c) CMEs can be divided according to their association with filament types: active region filaments, AR-CMEs; quiet-sun disappearing filaments, DSF-CMEs, and stealthy CMEs. The mean values of the peak magnetic field are $B_{max}$ = 18.5 nT for AR-ICMEs, 17.1 nT for DSF-ICMEs, and 11.6 nT for stealthy ICMEs.

d) A strong correlation is found between $B_{max}$ measured in the turbulent sheath (TS) and frontal region (FR), with $cc$ = 0.96. Somewhat weaker, but still well defined, are the correlations of $B_{max}$ values in TS and FR *vs.* those measured in the magnetic obstacle, MO ($cc$ = 0.87 and 0.86, respectively).

e) When taken separately, the TS/FR correlations for AR, DSF, and stealthy CME types show similar regression lines, with a correlation of $cc$ = 0.91, 0.98 and 0.95, respectively. The three subsets of data points are grouped in a compact cloud, resulting in the strong TS/FR correlation mentioned under item d). This indicates that there is a strong physical relationship between TS and FR, both consisting of compressed ambient solar wind plasma.

f) In the case of the TS/MO as well as FR/MO correlations, the AR, DSF, and stealthy types show a significantly different slope of the linear relationship for both the TS/MO and FR/MO correlations, and overall, the data-points are more scattered than in the TS/FR case. On the other hand, the individual correlations for the AR, DSF, and stealthy types are very strong for the TS/MO relationship ($cc$ = 0.97, 0.99 and 0.99, respectively), as well as for the FR/MC relationship ($cc$ = 0.85, 0.97 and 0.98 respectively).

g) Strong correlations are also found for the peak values of the flow speed, $v_{max}$ ($cc$ = 0.96, 0.93, and 0.94 for TS/FR, TS/MO, and FR/MO correlations, respectively).

h) In the case of the peak proton density, $Np_{max}$, a correlation is found only between the values measured in TS and FR ($cc$ = 0.78). The values of $Np_{max}$ in these two ICME segments are not correlated with those in the MO segment.

i) The peak proton thermal speed in TS is strongly correlated with that measured in FR ($cc$ = 0.95). The correlations of the TS and FR values *vs.* MO values are somewhat weaker ($cc$ = 0.77 and 0.84, respectively).

**Acknowledgements** We are grateful to the referee whose thoughtful comments helped to significantly improve the paper, as well as to the STEREO/SECCHI team (Goddard Space Flight Center, Naval Research Laboratory), the SOHO/LASCO team (Naval Research Laboratory, Max Planck Institute for Solar System Research), the SDO and Wind teams, for their open-data policy. B.V. and M.D. acknowledge funding from the EU H2020 grant agreement No. 824135 (SOLARNET) and support by the Croatian Science Foundation under the project 7549 (MSOC). A.M.V. acknowledges the Austrian Science Fund: FWF projects no. P24092-N16 and P27292-N20.